\documentclass[12pt]{iopart}
 
\usepackage[utf8]{inputenc}		
\usepackage{lmodern}           	
\usepackage{microtype} 
\usepackage[english]{babel}		

\usepackage{iopams}
\usepackage{amssymb}			
\usepackage{mathrsfs}			
\usepackage{siunitx}			

\usepackage{graphicx}		 	

\newcommand{\qpar }{q_\parallel}
\newcommand{\lamq }{\lambda_q}

\newcommand{\kappaperpn }{\kappa_{\perp,0}}

\newcommand{\kappaparn }{\kappa_{\parallel,0}}
\newcommand{\chiperp }{\chi_{\perp}}
\newcommand{\chiperpn }{\chi_{\perp,0}}
\newcommand{\chipar }{\chi_{\parallel}}
\newcommand{\chiparn }{\chi_{\parallel,0}}

\newcommand{\mw}{\mega \watt \per \square \metre}

\makeatletter
\def\maxwidth#1{%
\ifdim\linewidth>0.9\textwidth
    .7#1\textwidth
\else
    #1\linewidth
\fi}

\begin{document}
\title[Analytic 1D Approximation of the Divertor Broadening]{Analytic 1D Approximation of the Divertor Broadening \textit{S} in the Divertor Region for Conductive Heat Transport}
\author{D. Nille$^1$, B. Sieglin$^1$, T. Eich $^1$}
\address{$^1$ Max-Planck-Institute for Plasma Physics, Boltzmannstrasse 2, 85748 Garching, Germany}
\ead{Dirk.Nille@ipp.mpg.de}

\begin{abstract}
Topic is the divertor broadening $S$, being a result of perpendicular transport in the scrape-off layer and resulting in a better distribution of the power load onto the divertor target. Recent studies show a scaling of the divertor broadening with an inverse power law to the target temperature $T_t$, promising its reduction to be a way of distributing the power entering the divertor volume onto a large surface area.

It is shown that for pure conductive transport in the divertor region the suggested inverse power law scaling to $T_t$ is only valid for high target electron temperatures. For decreasing target temperatures ($T_t < 20\,$eV) the increase of $S$ stagnates and the conductive model results in a finite value of $S$ even for zero target temperature.
It is concluded that the target temperature is no valid parameter for a power law scaling, as it is not representative for the entire divertor volume. This is shown in simulations solving the 2D heat diffusion equation, which is used as reference for an analytic 1D model describing the divertor broadening along a field line solely by the ratio of the perpendicular to the parallel diffusivity.

By assuming the temperature dependence of these two quantities an integral form of $S$ is derived, relying only on the temperature distribution along the separatrix between X-point and target. Integration along the separatrix results in an approximation for $S$, being in agreement with the 2D simulations. This model is also applicable to scenarios including heat losses, e.g. due to radiation.

Convective transport can not be neglected for high recycling conditions, hence the derived expression for $S$ is not expected to hold. However, based on the non-vanishing parallel transport, respective the finite parallel transport time, the divertor broadening is expected to reach a finite value.
\end{abstract}

\maketitle

\section{Introduction}
\label{sec:introduction}

The description of the power load profile on the divertor targets relies on the knowledge of the heat transport in the scrape-off layer, especially in the divertor volume. There the perpendicular transport into the private flux region can significantly reduce the peak power load onto the divertor target. Recent studies show a scaling of the divertor broadening with an inverse power law to the target temperature $T_t$, promising its reduction to be a way of distributing the power entering the divertor volume onto a large surface area.

Simulations solving the 2D heat diffusion equation are used as reference for an analytic 1D model describing the divertor broadening along a field line solely by the ratio of the perpendicular to the parallel diffusivity. For parallel transport Spitzer-Härm conductivity is assumed and for the perpendicular transport an Spitzer-Härm like model.

By assuming the temperature dependence of these two quantities an integral form of $S$ is derived, relying only on the temperature distribution along the separatrix between X-point and target. Integration along the separatrix results in an approximation for $S$, being in agreement with the 2D simulations and allowing to discuss the benefit of a larger divertor spreading with respect to the effort needed to achieve lower target temperatures. This model is also applicable to scenarios including heat losses due to radiation.

Convective transport can not be neglected for high recycling conditions, hence the derived expression for $S$ is not expected to hold. However, based on the non-vanishing parallel transport, respective the finite parallel transport time, the divertor broadening is expected to reach a finite value.

Section \ref{sec:divertor_broadening} introduces a 1D function describing the target heat load in experiments, and its connection to the parallel and perpendicular heat diffusivities for pure conductive transport. Section \ref{sec:sec:diffusion_models} introduces basic diffusion models used to describe the heat diffusion in the scrape-off layer. A simple scaling for $S$ with the electron temperature is derived. Section \ref{sec:Simulation} explains the 2D simulations used as reference for the analytic analysis presented in Section \ref{sec:1D_analytic_analysis}. In Section \ref{sec:1D_analytic_analysis} an analytic expression for $S$ depending on the target temperature and target heat flux is derived. Neglecting other transport mechanisms an expression for a \textit{divertor averaged} temperature is given. The temperature dependence of $S$ obtained by the 1D model is compared to 2D simulations. Section \ref{sec:conclusion} summarises the preceding four chapters and concludes the findings of this analysis.

\section{The Divertor Broadening \textit{S}}
\label{sec:divertor_broadening}

\begin{figure}
\centering
\includegraphics[width=\maxwidth{}]{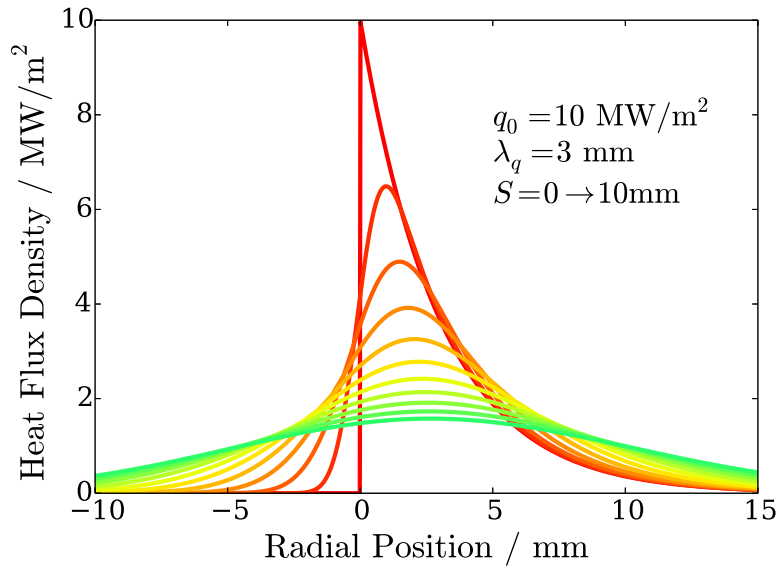}
\caption{Target heat flux profile steps for steps of 1\,mm in the the divertor broadening $S$, starting from the unperturbed X-point distribution in red up to $S = 10\,$mm in green.}
\label{fig:S_introduction}
\end{figure}

To describe the heat flux density profile on the divertor target, a model assuming only diffusive parallel and perpendicular electron conduction is used \cite{TEich:lamq}. All temperatures and densities in this paper refer to the electrons, being the dominant species for parallel diffusive transport for comparable ion and electron temperatures, as seen in the Braginski equations \cite{stangeby2000}. In the divertor volume the parallel heat diffusion time
\begin{equation}
\tau_\parallel = \frac{L^2}{\chipar}
\end{equation}
from the divertor entrance to the target is given by the connection length $L$, from the divertor entrance to the target, and the parallel diffusivity $\chipar$. The parallel diffusion time is equivalent to the perpendicular diffusion time for heat entering the divertor region. The perpendicular diffusion length is thus given by
\begin{equation}
S = \sqrt{\tau_\parallel \cdot \chiperp} = L \sqrt{\frac{\chiperp}{\chipar}}
\label{eq:S_tau}
\end{equation}
and is further called divertor spreading. Measurements of the heat flux profiles in AUG and JET are done by infrared thermography in target coordinates called $s$ with separatrix position $s_0$. Quantities following the magnetic field lines can be related to the upper midplane, to the radial coordinate called $x$, for comparison between different magnetic geometries and machines. The coordinates are correlated by the flux expansion $f_x$
\begin{equation}
x = \frac{s - s_0}{f_x}
\end{equation}
A power density profile given by a delta peak entering the divertor area is spread to a Gaussian of width $S$. The measure on the target is $S_{tar} = S\cdot f_x$. In this work $S$ refers to the divertor broadening mapped to the outer mid-plane if not marked otherwise.
\smallskip\\
The X-point heat flux density profile is described \cite{Makowski2012} by an exponential with peak value $q_0$ at the separatrix and decay length $\lamq$ at the midplane with the radial coordinate $x$:
\begin{equation}
q(x) = q_0 \cdot \exp\left(-\frac{x}{\lamq}\right) : x>0\;.
\end{equation}
\medskip
The target heat flux profile is described by the X-point profile convoluted with a Gaussian of width $S$, representing the broadening in the divertor region:
\begin{eqnarray}
q_\parallel (s) 
&=\frac{q_0}{2}\,\text{exp}\left( \left(\frac{S}{2\lamq}\right)^2 - \left(\frac{s -s_0}{f_x \lamq}\right)  \right)\cdot 
\mathrm{erfc}\left( \frac{S}{2 \lamq} - \frac{s-s_0}{f_x S}  \right)
\label{eq:q(s)}
\end{eqnarray}
Figure \ref{fig:S_introduction} shows the flattening of the heat flux density profile from the raw exponential in deep red -- starting at the strike point at $s_0=0$ -- up to a value of $S=10$\,mm in green in steps of $1$\,mm for $S$, keeping $\lamq$ and $q_0$ fixed.
\medskip\\
The peak heat flux $\hat{q}$ onto the target is used as design parameter and correlated to the integrated power profile of arbitrary shape $q(s)$ onto the target element by the integral power decay length 
\begin{equation}
\lambda_{int} \equiv \int \frac{q(s)}{\hat q} \rmd s
\label{eq:lamqint}
\end{equation}
The benefit of an additional divertor spreading $S$ on $\lambda_{int}$ compared to an exponential with decay length $\lamq$ is approximated \cite{Makowski2012} by:
\begin{equation}
\lambda_{int} \simeq \lamq + 1.64 \cdot S\;.
\label{eq:lamqint_scal}
\end{equation}
Studies predict a small value of $\lamq$ for future fusion relevant machines like ITER and DEMO compared to current machines like AUG and JET \cite{TEich:lamq}. Therefore the divertor spreading gains importance to meet the material limits of the divertor target with respect to the incoming heat flux density. Scaling laws for $S$ are available \cite{BSieglin_2013, BSieglin_2016a}, investigating $S$ for target electron temperatures above 20\,eV. Below this temperature, the increasing radiation prohibits IR measurements in AUG to deduce $S$. A study including simulations done in SOLPS \cite{Scarabosio_2014} shows a scaling inverse to the target electron temperature. The result with a fit to the SOLPS data is shown in figure \ref{fig:ST_scarabosio} for $S$ mapped upstream at ASDEX Upgrade (AUG) and JET. This will be further discussed in Section \ref{sec:conclusion}.
\medskip\\
The analysis on $S$ in this work is based on the transport model introduced in section \ref{sec:sec:diffusion_models} and is aimed to find a description of the divertor spreading with respect to the temperature distribution in the divertor volume. Intent is to quantify discrepancies in simulations and experiments for a better understanding of the transport mechanisms including a variety of particle and heat transport processes.

\section{Diffusion Models}
\label{sec:sec:diffusion_models}

Transport parallel to the magnetic field in the Scrape-off Layer (SOL) is described by Spitzer-Härm conduction with the conductivity
\begin{equation}
\kappa_\parallel = \kappaparn T^{5/2}.
\end{equation}
The factor $\kappaparn \simeq \SI{2000}{\watt\per\metre\electronvolt\tothe{-7/2}}$ is a valid approximation for machines like ASDEX Upgrade and JET for $Z_{eff} \approx 1.3$ \cite{stangeby2000}. The diffusivity is connected to the conductivity by the density \cite{venkanna2012}
\begin{equation}
\chi = \frac{\kappa}{\rho c_p} \,\propto\, \frac{\kappa}{n}\;.
\end{equation}
As only the ratio of the diffusivities is of interest, the factor representing the degrees of freedom of an electron in $c_p$ is not relevant. However, this can be important if several species contribute to the transport in parallel or perpendicular direction. To prevent confusion this factor is put into an alternative diffusivity
\begin{equation}
\tilde{\chi} = \frac{\kappa}{n}\;.
\end{equation}
Therefore the parallel diffusivity is given as
\begin{equation}
\tilde{\chi}_\parallel = \frac{\kappaparn}{n_e} \,T^{5/2}\;.
\end{equation}
The exponent for the temperature is referred to as $\beta \equiv 5/2$ in the further calculations.
Transport perpendicular to the magnetic field is characterised by a Spitzer-Härm like conductivity for this work, with the temperature dependence expressed by the exponent $\alpha$:
\begin{equation}
\tilde{\chi}_\perp = \frac{\kappaperpn}{n_e} T^{\alpha} 
\label{eq:chi_perp}
\end{equation}
Note that with this definition the density dependence in the ratio of the diffusivities cancels.
Bohm described the perpendicular diffusion coefficient \cite{Bohm_1949} in arc discharges as
\begin{equation}
\chi_{Bohm} = \frac{1}{16}\frac{T}{eB}
\end{equation}
being not dependent on $n_e$ but on $B$, as suggested by experiments \cite{BSieglin_2016a}. Neglecting the dependency on the total magnetic field, we find a perpendicular diffusion coefficient scaling linear with $T$. Using Bohm diffusion for perpendicular transport, the density dependence of the parallel diffusivity remains in the ratio of the diffusivities. To eliminate this explicit dependency, the ideal gas law $p = nT \rightarrow \frac{1}{n} = \frac{T}{p}$ can be used to add the inverse density dependence to a perpendicular diffusivity scaling with $T^{\alpha'}$. For simplicity the factor $k_B$ is not explicitly written.
\begin{equation}
\frac{1}{p}\cdot T^{\alpha'} = \frac{1}{p}\cdot T^{\alpha'}\,\cdot\,\frac{1}{n}\,\frac{p}{T} = \frac{T^{\alpha'-1}}{n} = \frac{T^\alpha}{n}\;
\end{equation}
with the new exponent
\begin{equation}
\alpha' = \alpha + 1
\end{equation}
and the corresponding diffusivity rewritten from equation \eref{eq:chi_perp}
\begin{equation}
\tilde{\chi}_\perp = \frac{\kappaperpn}{p} T^{\alpha+1}\;. 
\end{equation}
 Thus the density dependence is substituted for the pressure and subsequently from \eref{eq:S_tau} $S$ scales with
\begin{equation}
S\,\propto\,\sqrt{\frac{\chiperp'}{\chipar}} = \sqrt{\frac{\kappaperpn}{\kappaparn}} \,\cdot\,\sqrt{ \frac{T^{\alpha}}{T^{\beta}}} \,\propto\, T^{\frac{\alpha-\beta}{2}} = \sqrt{p}\cdot T^{\frac{\alpha'-\beta}{2}}\;.
\end{equation}
Including the density implicitly into the temperature is correct for pressure conservation and an analysis along a single field line. For a non-constant pressure $S$ scales with the inverse square root of the pressure. Recent studies suggest a scaling of $S$ with about the inverse square root of the divertor density \cite{BSieglin_2016a}. This corresponds to a weak or no density dependence of $\chi_\perp$. The formulation where the density is treated implicitly in the temperature assuming pressure conservation is used in the 1D Model. This corresponds to $\alpha'=0$ respective $\alpha = 1$. 
\medskip\\
Other numerical tools used to study the heat and particle transport in the SOL are using similar, but not necessarily the same expressions and approximations for the diffusion coefficients. SOLPS for example assumes a constant perpendicular diffusivity throughout the entire SOL. This corresponds to $\alpha = 1$ -- respective $\alpha'=0$ -- in the model introduced in this section.

\section{Simulation}
\label{sec:Simulation}

As reference for the 1D analysis a 2D model in slab geometry is used, solving the heat diffusion equation in the SOL. For parallel transport Spitzer-Härm conduction is assumed. For perpendicular transport a Spitzer-Härm like diffusivity as described in section \ref{sec:sec:diffusion_models} with a fixed temperature exponent $\alpha$ and inverse density dependence is used.
\medskip\\
The heat diffusion equation is solved using the heat potential
\begin{equation}
u(\kappa) = \int\limits_{0}^{T} \kappa(T') dT'
\end{equation}
leading to the linear differential equation
\begin{equation}
\frac{dT}{dt} = \chi\Delta u
\end{equation}
instead of the non-linear second order partial differential equation
\begin{equation}
\frac{dT}{dt} = \frac{1}{n}\nabla\left(\kappa(T) \nabla T\right)\;.
\end{equation}
An alternate direction implicit Crank-Nicolson scheme is implemented to solve the heat diffusion equation \cite{numericalrecipes}.
\medskip\\
In the simulation and analytic analysis the pressure is assumed to be constant along field lines. In addition, to be independent of actual pressure distributions, the simulation uses a homogeneous pressure distribution in the divertor volume. For simplicity and better comparability the target temperature in the simulation used for comparison in this paper is set to a constant value at the target surface. Setting the temperature distribution according to the sheath model, including the incident heat flux and the density distribution in front of the target, shows a minor impact on the resulting heat flux profiles. $S$ increases up to 25\% for the studied cases, setting a temperature distribution according to the sheath model instead of the peak temperature for the entire tile. Therefore the simulation results are a pessimistic approach for the divertor broadening for increasing target temperature.

\subsection{Geometric Configurations}
\label{sec:sim_geo}

\begin{figure}
\centering
a)\\
\includegraphics[width=.5\textwidth]{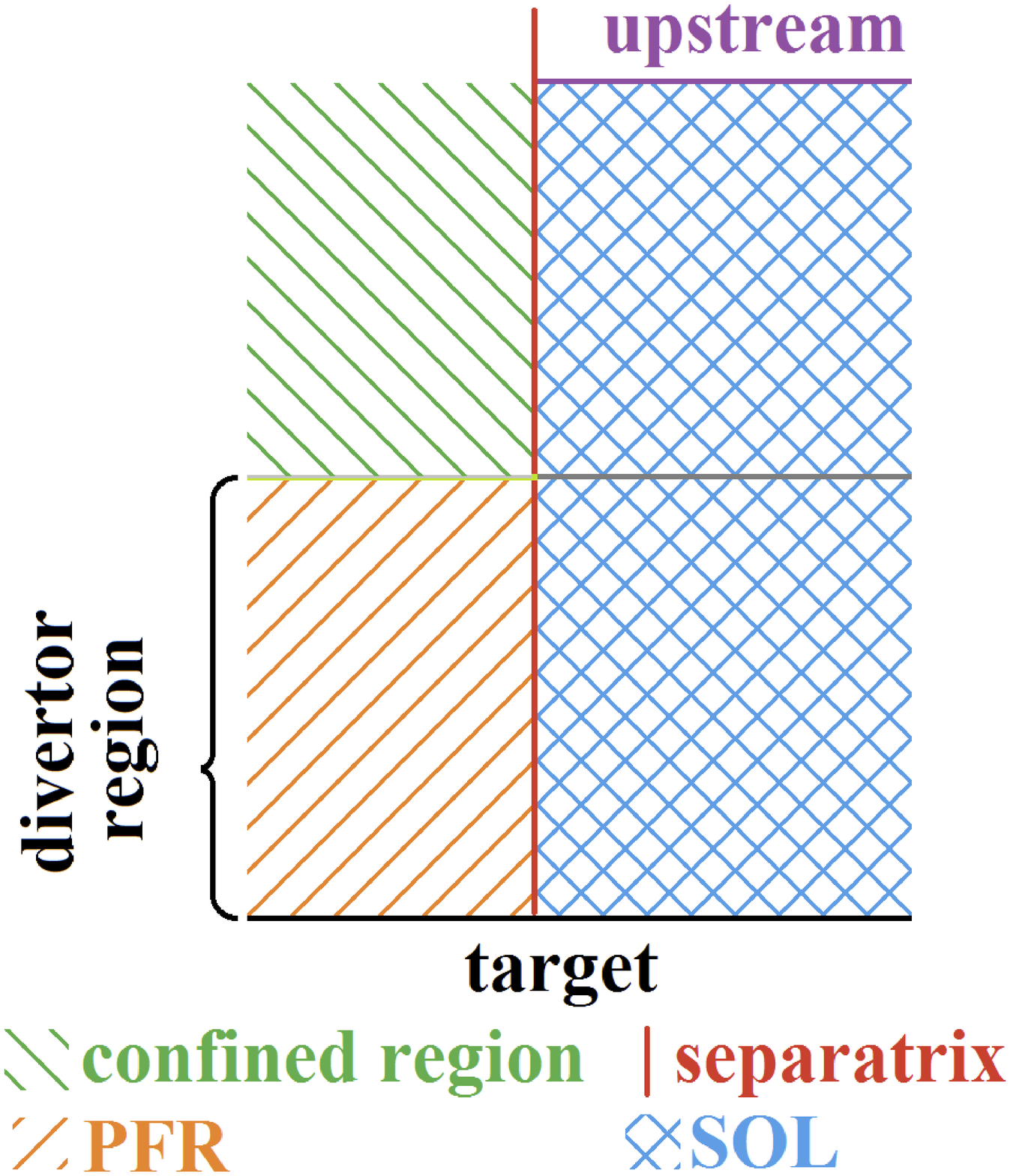}
\\b)\\
\includegraphics[width=\maxwidth{}]{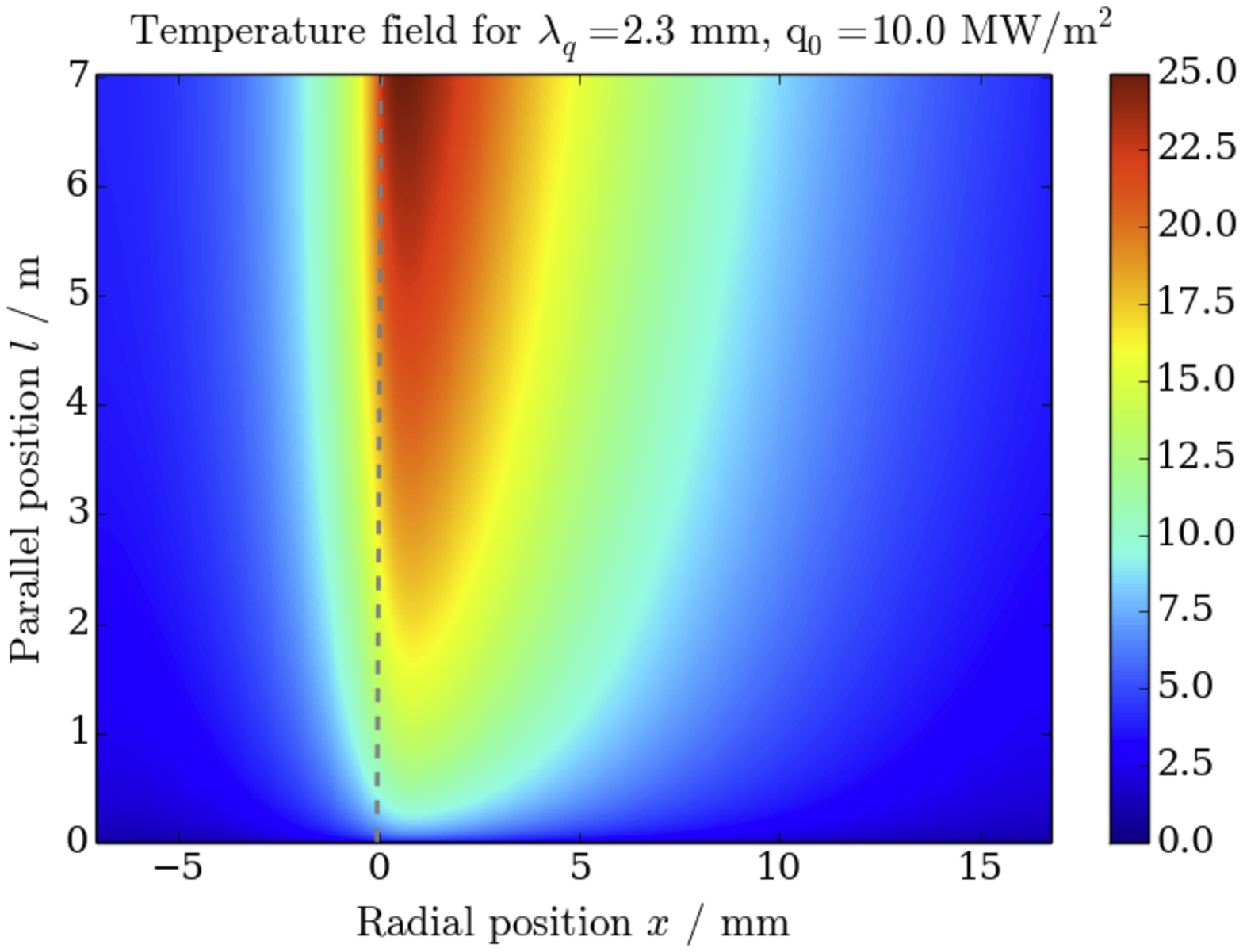}
\caption{a) Sketch of the slab geometry. b) Example of a temperature field resulting from a simulation, restricted to the divertor region.}
\label{fig:simulation_geometry}
\end{figure}

The simulation is able to solve the 2D heat diffusion equation in various topologies. Figure \ref{fig:simulation_geometry} shows a comparison between a) the structure of the slab geometry and b) the computational grid with a steady state temperature field. A column of the array represents the path along the magnetic field, a row represents the direction perpendicular to it. The simulation can include the SOL above the X-point, but for studying $S$ for given X-point conditions only the divertor region is included. This is called the divertor configuration. The divertor target is at the bottom, the divertor entrance at the top and the separatrix is marked with vertical lines. The confined region is excluded from the calculations, as the diffusion models do not describe the transport in this region. Upstream -- respective at the divertor entrance for the divertor configuration -- an exponentially decaying parallel heat flux density is set as boundary condition. The upper boundary of the PFR represents the connection to the PFR of the other divertor. No heat exchange is allowed at this boundary. Downstream the temperature is used as boundary condition. For the example shown the length of the divertor leg is set to 7\,m -- based on ASDEX Upgrade. A constant connection length is assumed.
\medskip\\
The presented results stem from simulations restricted to the divertor volume. The benefit is that at the entrance the power fall off width $\lamq$ and the peak heat flux $q_0$ are known, as the heat flux density is used as boundary condition, without being changed by perpendicular transport above the divertor entrance. These two geometric configurations show only minor differences in the resulting target heat flux profiles, for comparable $\lamq$ and $q_0$ at the divertor entrance, with the divertor configuration being better suited to study the divertor broadening.

\section{1D Analytic Analysis on \textit{S}}
\label{sec:1D_analytic_analysis}

\begin{figure}[htb]
  \center
  \includegraphics[width=\maxwidth{}]{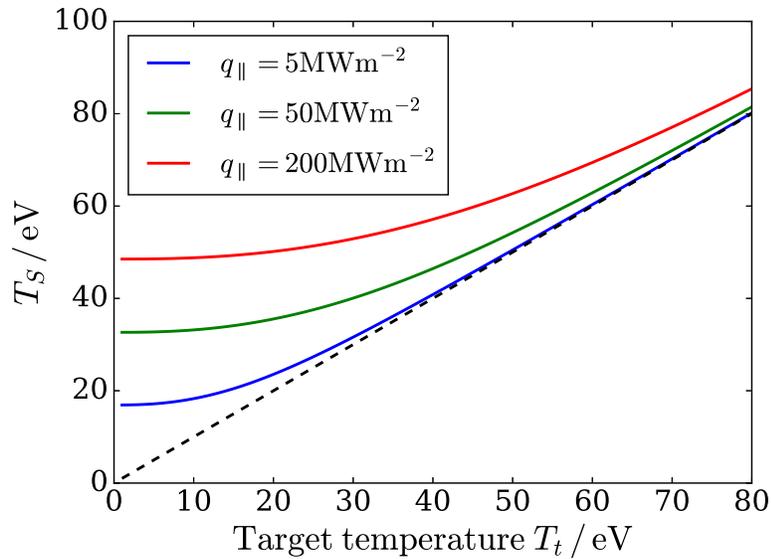}
  \caption{Effective divertor spreading temperature $T_S$ for varying target temperature $T_t$ for different parallel heat flux densities.}
  \label{fig:TS_Tt}
\end{figure}

\begin{figure}[htb]
  \center
  \includegraphics[width=\maxwidth{}]{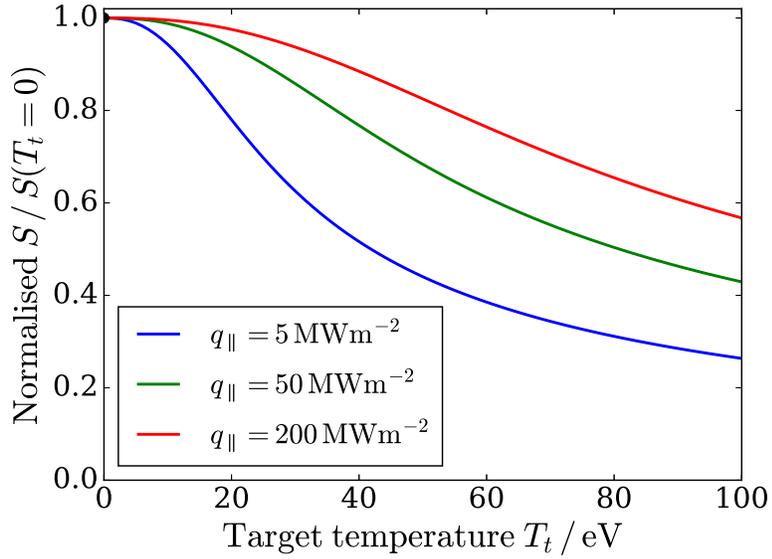}
  \caption{Divertor spreading normalised to $S(T_t=0)$ for varying target temperature and three different parallel heat flux densities.}
  \label{fig:S_Tt}
\end{figure}

Interpreting equation \eref{eq:S_tau}
\begin{equation}
S \approx L \sqrt{\frac{\chiperp}{\chipar}}\;,
\label{eq:S_tau2}
\end{equation}
in which the diffusivities resemble averaged values, as integral parallel to the magnetic field leads to
\begin{equation}
S = \int_0^L \sqrt{\frac{\chiperp}{\chipar}} \rmd l = \sqrt{\frac{\chiperpn}{\chiparn}} \int_0^L  T(l)^{\frac{\alpha-\beta}{2}} \rmd l\;.
\label{eq:Sint1}
\end{equation}
Assuming a constant parallel heat flux $q_\parallel$ the temperature profile described by the two point model is
\begin{equation}
T(l) = \left(T_t^{\beta+1} + (\beta+1)\frac{\qpar\cdot l}{\kappaparn} \right) ^{1/(\beta+1)}
\label{eq:TPM}
\end{equation}
depending on the target temperature $T_t$ and distance $l$ from the target. For $T_t = 0$ the X-point temperature ($l=L$) is defined as
\begin{equation}
T_{X,0} = \left((\beta+1)\frac{\qpar\cdot L}{\kappaparn} \right) ^{1/(\beta+1)}
\end{equation}
With this expression equation \eref{eq:TPM} is rewritten as
\begin{equation}
T(l) = \left(T_t^{\beta+1} +T_{X,0}^{\beta+1}\frac{l}{L} \right) ^{1/(\beta+1)}\;.
\end{equation}
Equation \eref{eq:Sint1} is expressed by using the derived term for $T(l)$ containing $l$ explicitly in the integral:
\begin{equation}
S = \sqrt{\frac{\chiperpn}{\chiparn}} \cdot \int_0^L \left(T_t^{\beta+1} + T_{X,0}^{\beta+1} \cdot\frac{l}{L}\right) ^{\frac{\alpha - \beta}{2(\beta + 1)}}\, \rmd l
\end{equation}
The exponent of the integrand is negative for a stronger temperature dependence of the parallel transport -- given $\beta > \alpha$ -- and is in the range of -0.5 to -0.2 for $ -1 \le \alpha \le +1$.
The solution to the integral is
\begin{eqnarray}
S =& L\cdot\sqrt{\frac{\chiperpn}{\chiparn}}\cdot \frac{2(\beta+1)}{\alpha + \beta + 2} \,\cdot
\frac{\left( T_t^{\beta + 1} + T_{X,0}^{\beta + 1} \right)^{\frac{\alpha + \beta + 2}{2(\beta+1)}} - T_t^{\frac{\alpha + \beta + 2}{2}}}{T_{X,0}^{\beta +1}}\;.
\label{eq:Sint}
\end{eqnarray}
It expresses S by $T_{X,0}$ -- given by $q_\parallel$ -- and $T_t$, both being measurable quantities in the experiment. The difference of the exponents of the temperature dependencies $\beta - \alpha$ is positive for experimental relevant SOL transport.
\medskip\\
In the case $T_t = 0$ the result is
\begin{equation}
S = L \sqrt{\frac{\chiperpn}{\chiparn}}\, T_{X,0}^{-\frac{\beta - \alpha}{2}} \cdot \frac{2(\beta + 1)}{\alpha + \beta +2}\;.
\label{eq:S_Ttn}
\end{equation}
This finite value is scaling inverse with the temperature at the X-point and with the square root of the ratio of the temperature independent diffusivity factors.
In equation \eref{eq:S_Ttn}
\begin{equation}
\frac{2(\beta + 1)}{\alpha + \beta +2} = \mathrm{const}
\end{equation}
is identified as a constant for given transport models, describing the effective divertor temperature
\begin{equation}
T_S = T_{X,0} \cdot \left(\frac{2(\beta + 1)}{\alpha + \beta +2}\right)^{-\frac{2}{\beta - \alpha}}
\end{equation}
for 
\begin{equation}
S = L \sqrt{\frac{\chiperpn}{\chiparn}}\,T_S^{-\frac{\beta - \alpha}{2}}\;.
\label{eq:S_TS}
\end{equation}
Note the explicit linear dependence on the connection length to the X-point and the inverse scaling with the averaged temperature. As the connection length is increased, the effective temperature $T_S$ increases. As a result the divertor spreading is increasing less than linear with the divertor length. To significantly increase $S$, the effective temperature has to be reduced, which can be achieved by increasing the divertor volume.
\medskip\\
For a non-constant $q_\parallel(l)$, for example lowered due to the perpendicular diffusion, the integral can be iterated numerically to find $S$ for given $q_0$ and $T_t$. This approach also allows to take losses, e.g. due to radiation, into account. 
\medskip\\
Figure \ref{fig:TS_Tt} shows the resulting value for $T_S$ for given $T_t$ for connection length $L=7$\,m, $\alpha=1$ and three parallel heat flux densities. For this comparison the heat flux density is assumed to be constantly $q_0$, to be independent of the actual spreading, giving an upper boundary. Note that the target heat flux densities are obtained from the parallel heat flux density in the plasma approaching the target by taking the field line inclination angle and the geometric flux expansion into account. The higher the target temperature, the closer is the effective temperature to the target value, as the parallel temperature gradient decreases for increasing temperature for the same heat flux density. 
\\
Figure \ref{fig:S_Tt} shows the divertor spreading normalised to 1 for $T_t=0$. The decrease of $S$ depends on the parallel heat flux density, which is like in figure \ref{fig:TS_Tt} kept constant. This graph shows, that the analysis of $S$ with the target temperature as reference is expected to depend on the parallel heat flux density $q_0$.

\subsection{Approaches to $q(l)$}

Taking the divertor spreading in the 1D model into account, the question is how to calculate $q(l)$ arises. $q_0$ by definition is the peak heat flux at the divertor entrance. A decrease of the parallel heat flux density along the divertor volume reduces the parallel temperature gradient. A pessimistic approach is to use the peak heat flux according to the integral decay length \eref{eq:lamqint} as $S$ increases along the divertor leg:
\begin{equation}
q(l) = q_0 \cdot \frac{\lamq}{\lamq + 1.64 S(l)}\;.
\end{equation}
This approach is pessimistic, as the radial position of the peak heat flux is moving into the SOL as the heat flux profile degrades by perpendicular transport. Using the peak heat flux independent of its radial position leads to the steepest parallel temperature gradient and therefore to an upper limit of the temperature evolution along the field lines near the separatrix. 
\medskip\\
A less pessimistic approach is given, using the temperature profile along a single field line, located in the SOL. Therefore equation \eref{eq:q(s)} can be evaluated with $S(l)$. The issue is the dependence of the result on the chosen distance to the separatrix. For values much smaller than the divertor broadening $x \ll S$ the parallel heat flux density drops quickly after the X-point, due to the perpendicular transport into the PFR. For $x \approx S$ the parallel profile $q(l)$ approaches the shape of the pessimistic method described before this method, but stays below $q_0$ at the divertor entrance. Due to the drawback of the shape dependence it is not feasible to use $q$ at a fixed radial distance to the separatrix.

\subsection{Comparison to 2D Calculation and Experiment}

\begin{figure}[htb]
  \center
  \includegraphics[width=\maxwidth{}]{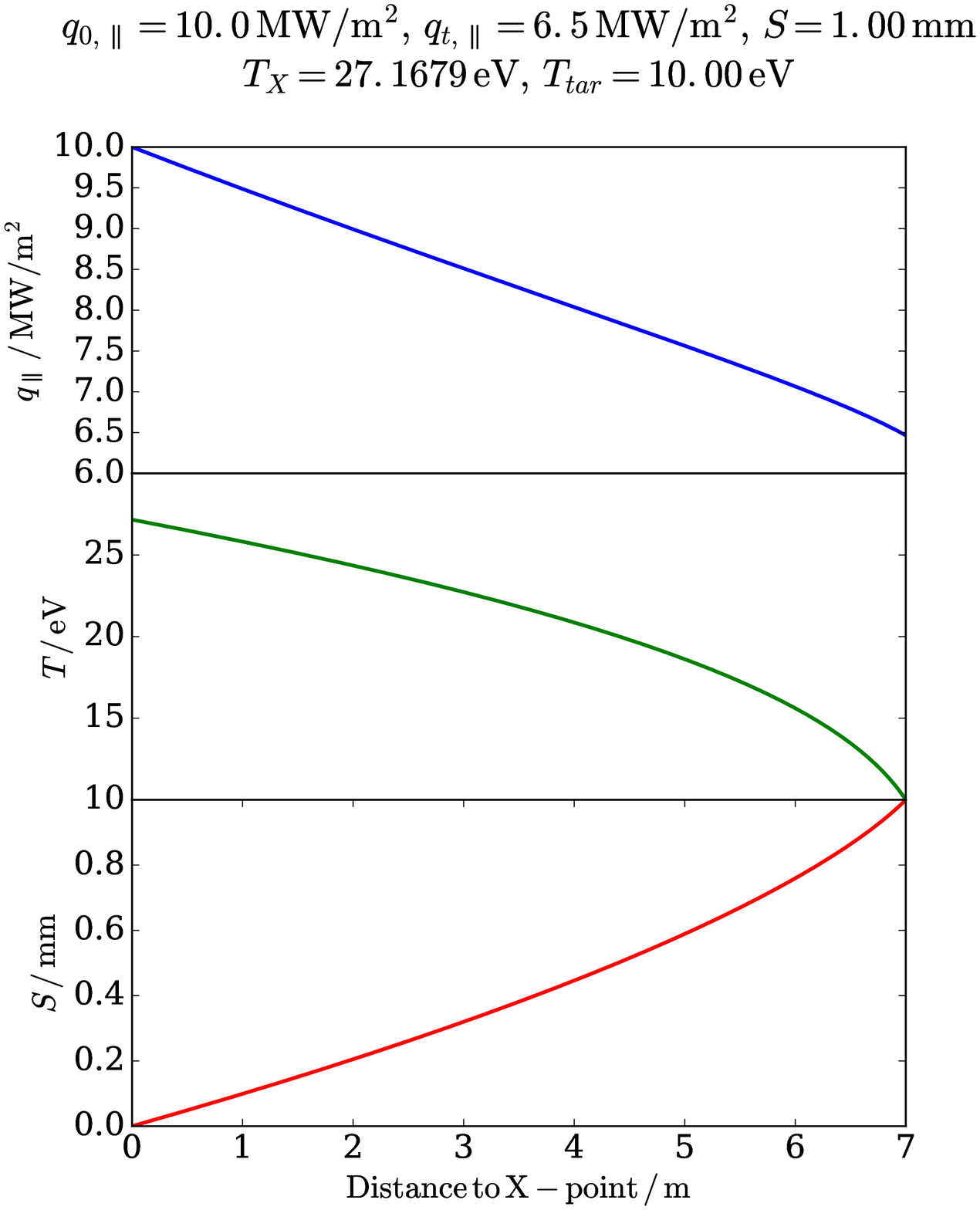}
  \caption{Parallel profiles of parallel heat flux density $q_\parallel$, electron temperature $T$ and divertor broadening $S$. $\alpha = 0$.}
  \label{fig:spreading1}
\end{figure}

\begin{figure}[htb]
  \center
  \includegraphics[width=\maxwidth{}]{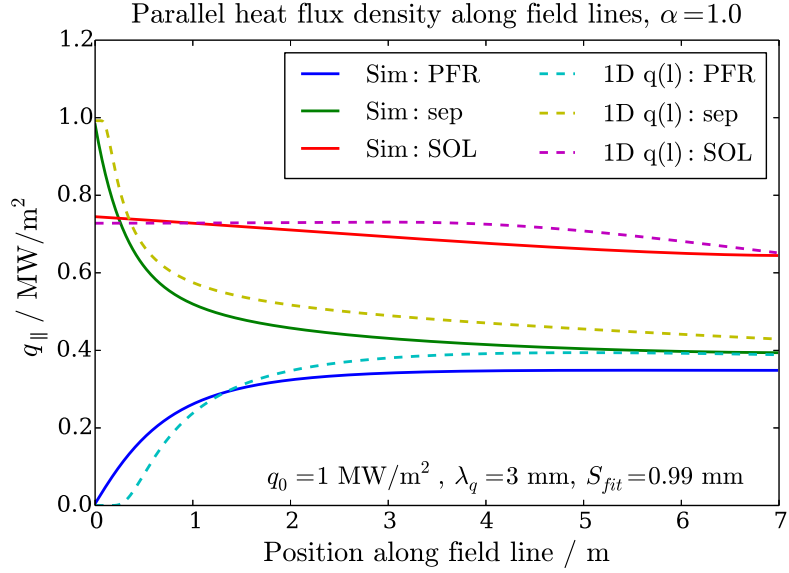}
  \caption{Heat flux density along three field lines from the 2D simulation in comparison to the 1D model assuming linear dependence of $S$ to $l$. Simulation results are shown as solid lines, 1D approximations as dashed lines.}
  \label{fig:qpar_sim_1D}
\end{figure}

Figure \ref{fig:spreading1} shows the evolution of the peak parallel heat flux, the temperature and the divertor broadening $S$ along the divertor volume. Boundaries are $q_0 = \SI{10}{\mw}$, $S = 1$\,mm and $T_t=10$\,eV. The gradient of $S$ is highest near the target, as the falling temperature reduces the parallel transport stronger than the perpendicular transport. Assuming a linear increase of $S\,\propto\,l$ from the X-point to the target is reasonable for models taking the heat flux density in the divertor volume into account, without treating the broadening mechanism. Figure \ref{fig:qpar_sim_1D} shows a comparison between the parallel heat flux density close to the separatrix in the SOL (sep), in the SOL and in the PFR. The deviation of the profiles at the target is due to a mismatch between the target heat flux profile and the 1D model for the target heat flux density profile, see equation \eref{eq:q(s)}. It is known that a regression of this function to data overestimates the heat flux density close to the separatrix and underestimates it further in the PFR when compared to experiments, which is reproduced in the 2D simulation.

\begin{figure}[htb]
  \center
  \includegraphics[width=\maxwidth{}]{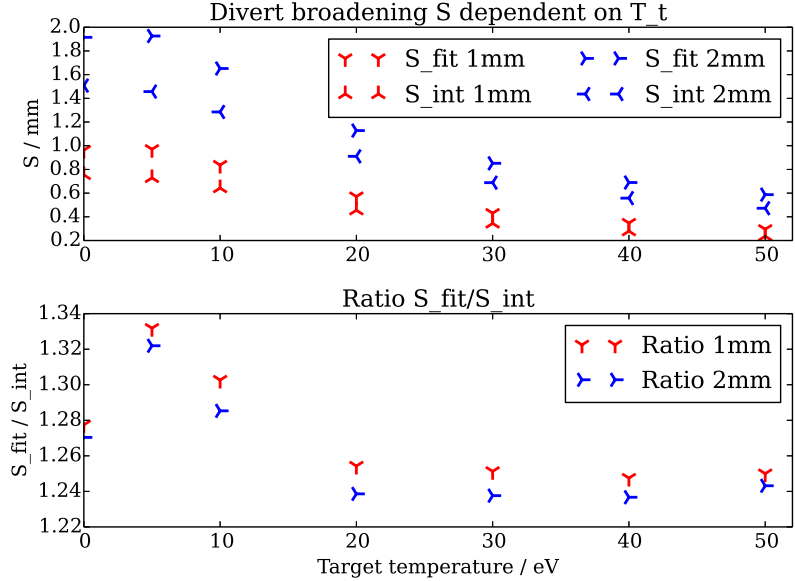}
  \caption{Ratio of $S$ from 2D Simulation to 1D integral for constant parallel heat flux density $q = q_0 =$. $\alpha = 0$. Note that the same ratio for the diffusivities is used.}
  \label{fig:spreading2}
\end{figure}

\begin{figure}[htb]
  \center
  \includegraphics[width=\maxwidth{}]{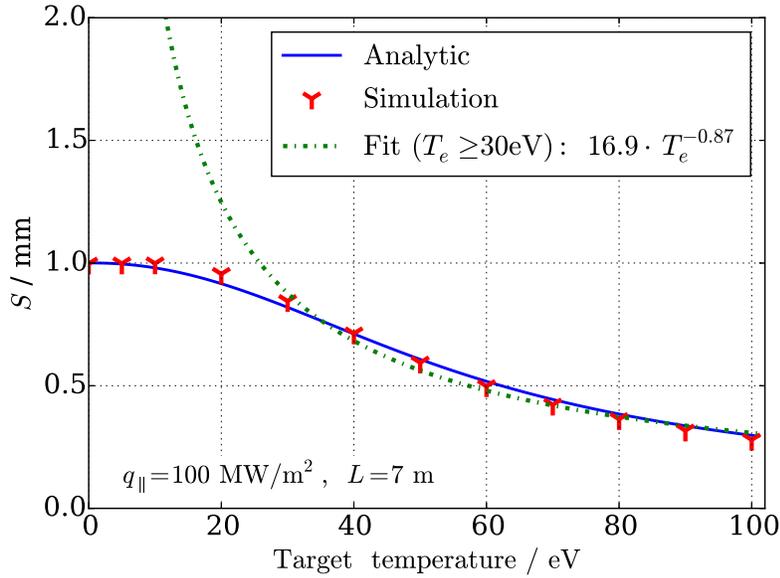}
  \caption{Best fit of inverse power law to describe $S$ based on $T_t$ for $T_t \ge 30$\,eV. $\alpha = 0$.}
  \label{fig:ST_fit}
\end{figure}

\begin{figure}[htb]
  \center
  \includegraphics[width=\maxwidth{}]{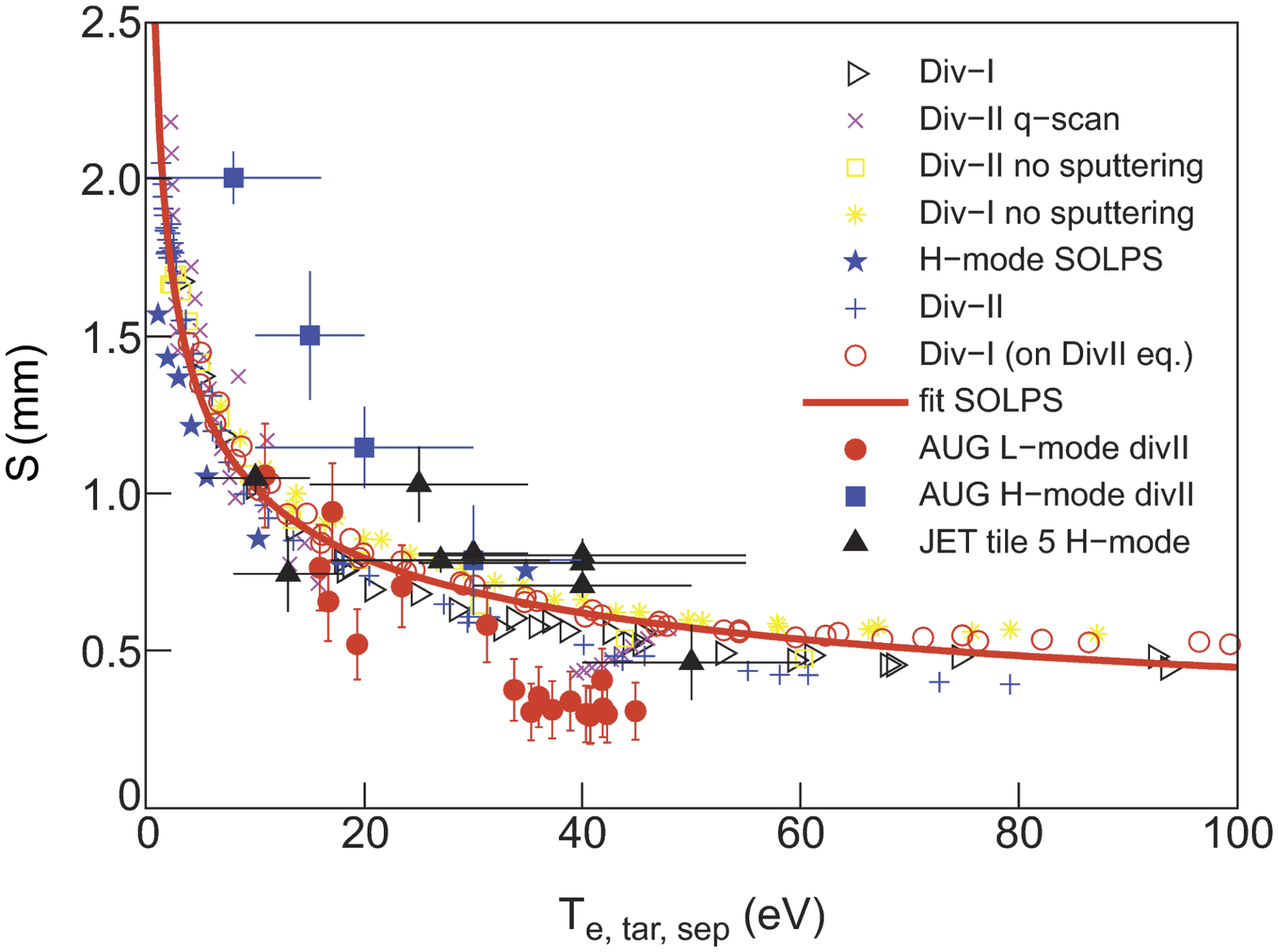}
  \caption{$S$ as function of target electron temperature at the separatrix $T_{e,tar,sep}$, based on \cite{Scarabosio_2014}.}
  \label{fig:ST_scarabosio}
\end{figure}

A comparison of the absolute divertor spreading $S$ between 1D model and 2D calculation is shown in figure \ref{fig:spreading2} for diffusivities set to yield two different $S$ of 1 and 2\,mm for $T_t = 0$ in the 2D simulation. The upper graph shows the absolute value of $S$ for varying target temperature. The lower graph shows the ratio of the fitted 2D data to the 1D integral result. The integral \eref{eq:Sint1} underestimates $S$ at around 20\% compared to the 2D calculation, but agrees with the trend. Figure \ref{fig:ST_fit} shows the trends of 2D and 1D results for $S$ with the pre factor being corrected by regression for $\alpha = 0$.
Figure \ref{fig:ST_scarabosio} shows experimental and simulated values from \cite{Scarabosio_2014} for $S = \frac{S_{tar}}{f_x}$ based on the value $S_{tar}$ measured on the target. Shown are measured data from JET and ASDEX Upgrade. The regression uses data deduced from the SOLPS simulations. The best fit for an inverse power law is given by
\begin{equation}
{S} = (2.3\pm0.2) T_e^{-0.36\pm0.03}\;.
\end{equation}
In contrast a fit for temperatures above 30\,eV of the analytic function and the corresponding 2D simulation yields an exponent of -0.87, shown in figure \ref{fig:ST_fit}. The pre factor is not of interest, as a ratio of diffusivities leading to $S=1$\,mm for $T_t=0$ is used. For this case of $\alpha = 0$ the fit tends towards an exponent of -1.25 by including higher target temperatures, as expected from equation \eref{eq:S_TS} for a flat temperature profile with $T(l) \approx T_t$ from target to X-point.
\medskip\\
It should be distinguished between an energy conserving broadening due to perpendicular transport and a flattening of the peak heat flux by losses due to radiation, charge exchange etc for deducing transport parameters from target heat flux profiles. These processes can lead to an overestimation of $S$ in simulations -- with respect to perpendicular broadening -- with $q_0$ being underestimated.

\section{Conclusions}
\label{sec:conclusion}

An approximation for $S$ is found by integrating the temperature profile along the separatrix between X-point and divertor target. While a power law scaling for $T_t$ is valid for high target temperatures, this is not true for low target temperatures due to the strong parallel temperature gradient. This approach implies, that it is not enough to reduce the plasma temperature close to the target for larger machines like ITER and DEMO to achieve the required low target heat flux, but the temperature has to be lowered in a large volume in front of the target. This could be achieved by e.g. volume radiation in the divertor volume.
\medskip\\
An important consequence is, that an increase of the length of the divertor leg -- in which conduction is the dominant transport parameter -- will increase the divertor spreading less than linear, due to the strong temperature gradient in front of the target and therefore high temperature near the X-point with low contribution to $S$. Additional connection length, however, can be used to decrease the parallel heat flux density in the divertor volume by radiation and dissipative processes, so that the temperature gradient in front of the target is lowered, which leads to an increase of $S$.
\medskip\\
The presented 1D analysis neglects processes like convective transport, drifts and radiation, which limits the predictive capability of the model, as these are known to have an influence on the target heat flux profile. The 2D simulation used for comparison assumes a constant pressure in the divertor volume and a homogeneous target temperature for better comparability. Both neglect the influence of neutral particles, known to be important for detachment, reducing heat and particle flux to the target.
\medskip\\
Therefore this model delivers an approximation for attached conditions, with conduction being the dominating transport in the divertor volume.

\section{References}
\bibliography{references.bib}

\end{document}